\documentclass[conference]{IEEEtran}
\IEEEoverridecommandlockouts
\usepackage{cite}
\usepackage{amsmath,amssymb,amsfonts}
\usepackage{algorithmic}
\usepackage{graphicx}
\usepackage{textcomp}
\usepackage{xcolor}
\usepackage[utf8]{inputenc}
\usepackage[T1]{fontenc}
\def\BibTeX{{\rm B\kern-.05em{\sc i\kern-.025em b}\kern-.08em
    T\kern-.1667em\lower.7ex\hbox{E}\kern-.125emX}}
\begin{document}

\title{IR-UWB Radar-Based Contactless Silent Speech Recognition with Attention-Enhanced Temporal Convolutional Networks
\thanks{This work was supported by the National Research Foundation of Korea (NRF) grant funded by the Korea government (MSIT) (NRF-2021R1F1A1062958).
}
}

\author{\IEEEauthorblockN{Sunghwa Lee${}^{*}$} 
\IEEEauthorblockA{\textit{School of Integrated Technology} \\
\textit{Yonsei University}\\
Incheon, Korea \\
sunghwa.lee@yonsei.ac.kr}
{\small${}^{*}$ Corresponding author}
\and
\IEEEauthorblockN{Jaewon Yu} 
\IEEEauthorblockA{\textit{School of Integrated Technology} \\
\textit{Yonsei University}\\
Incheon, Korea \\
jaewon.yu@yonsei.ac.kr}
}

\maketitle

\begin{abstract}

Silent speech recognition (SSR) is a technology that recognizes speech content from non-acoustic speech-related biosignals.
This paper utilizes an attention-enhanced temporal convolutional network architecture for contactless IR-UWB radar-based SSR, leveraging deep learning to learn discriminative representations directly from minimally processed radar signals.
The architecture integrates temporal convolutions with self-attention and squeeze-and-excitation mechanisms to capture articulatory patterns.
Evaluated on a 50-word recognition task using leave-one-session-out cross-validation, our approach achieves an average test accuracy of 91.1\% compared to 74.0\% for the conventional hand-crafted feature method, demonstrating significant improvement through end-to-end learning.

\end{abstract}

\begin{IEEEkeywords}
Contactless silent speech recognition, impulse radio ultra-wideband (IR-UWB) radar, temporal convolutional networks, self-attention, squeeze-and-excitation
\end{IEEEkeywords}

\section{Introduction}

Silent speech recognition (SSR) deciphers speech content from non-acoustic speech-related biosignals.
SSR offers an alternative or a complement to existing acoustic-based automatic speech recognition.
This enables communication in environments that are either too noisy (e.g., construction sites, factories, crowded areas) or require silence (e.g., libraries, operating rooms), as well as for individuals unable to produce acoustic speech.

A range of sensing methods have been employed to facilitate SSR  \cite{Denby2010, Schultz2017, Gonzalez2020:SSIreview, Lee2021:SSIreview}.
Some sensing techniques, including electromagnetic articulography \cite{Heracleous2011:EMA, Kim2017} and permanent magnetic articulography \cite{Gilbert2010, Gonzalez2016}, acquire nonacoustic speech-related biosignals through intraoral and perioral instrumentation.
They are effective in capturing detailed articulatory movements, but the need for intraoral sensor placement can limit user comfort and usability.

Surface-contact technologies, exemplified by surface electromyography \cite{Meltzner2017, Zhu2021:sEMG, Deng2023:sEMG, Tan2023:sEMG, Huang2024:sEMG}, and contactless methods, such as vision-based sensing \cite{Sheng2021:image, Lopez2022:image, Haq2022:image, Chen2024:lipreading, Tan2025:lipreading}, offer a balance between usability and signal acquisition.
Despite their advantages over intraoral approaches in terms of usability, these external methods still face significant limitations: contact-based systems require sensor attachment and suffer from placement inconsistencies \cite{Prorokovic2019:sEMG}, while vision-based contactless techniques struggle with detecting hidden articulators and maintaining consistent performance across environments. 

Impulse-radio ultra-wideband (IR-UWB) radar \cite{Lee17:IR-UWB, Lee18:Simulation, Lee18:IR-UWB, Shin17:Autonomous} holds a lot of promise in SSR applications.
Unlike surface-contact technologies that require sensor attachment and suffer from placement inconsistencies, IR-UWB radar enables completely contactless operation while maintaining accurate speech movement capture through high range resolution.
IR-UWB radar also preserves user privacy and maintains reliable performance under varying environmental conditions.
These characteristics establish IR-UWB radar as particularly well suited for developing practical SSR systems that function consistently in everyday applications.

Recent IR-UWB-based SSR research includes recognizing words \cite{Shin16:Towards, Lee19:Word} and more comprehensive recognition of vowels, consonants, words, and phrases \cite{Lee23:IR-UWB}, demonstrating the potential of IR-UWB radar for SSR.
Although these studies attained notable results, they primarily relied on hand-crafted features for speech recognition.

Meanwhile, modern deep learning approaches have shown that learnable feature extraction can replace manually designed features across various domains.
Temporal convolutional networks (TCNs) \cite{Bai2018:TCN} have demonstrated effectiveness in sequence modeling tasks by capturing long-range dependencies through dilated convolutions while maintaining computational efficiency.
The attention mechanism, particularly self-attention introduced in Transformer architectures \cite{Vaswani2017:Attention}, has revolutionized sequence processing by enabling models to focus on relevant parts of the input dynamically.
Furthermore, channel attention mechanisms such as squeeze-and-excitation (SE) blocks \cite{Hu2018:SE} have shown significant improvements in feature representation learning by adaptively recalibrating channel-wise feature responses.
Our work leverages these advances, combining TCNs with attention mechanisms for IR-UWB-based silent speech recognition.

Specifically, we utilize an attention-enhanced temporal convolutional network architecture for IR-UWB radar-based SSR.
This architecture employs temporal convolutional layers as primary feature extractors, capturing local patterns and hierarchical representations from radar signals.
These convolutional components effectively process the complex temporal patterns present in radar data that capture articulatory movements.

The self-attention mechanism enhances feature extraction by computing relationships between all temporal positions in the sequence, allowing the model to capture long-range dependencies and focus on relevant articulatory patterns regardless of their temporal distance.
Additionally, the squeeze-and-excitation blocks provide channel-wise attention, automatically emphasizing informative feature channels while suppressing less useful ones.
Together, these attention mechanisms address a key challenge in radar-based SSR: identifying subtle articulatory movements from complex received radar returns.

The integration of these complementary components---temporal convolutions for hierarchical feature extraction, self-attention for temporal dependencies, and SE blocks for channel recalibration---creates a robust architecture that enables IR-UWB radar-based SSR with minimal signal processing.
We evaluate our approach on a 50-word recognition task, demonstrating superior performance compared to the conventional approach based on hand-crafted features.
Unlike previous approaches that rely on hand-crafted features followed by separate classification models, our end-to-end approach learns discriminative representations from radar signals with minimal preprocessing.

\section{Methodology}

\subsection{Data Collection}

We adopted a hardware testbed for data collection of radar-based SSR.
This testbed was originally developed in \cite{Lee23:IR-UWB} with two radar modules (NVA-R661, Novelda)---one positioned in front of the lips and another under the cheek---and the study demonstrated that the front-lip placement achieves superior SSR performance.
Therefore, we used only the single radar module positioned in front of the participant's lips, with the antenna directed toward the lip area.

For data acquisition, we employed the MATLAB-based GUI system from \cite{Lee23:IR-UWB} that allows participants to self-administer the collection process.
This system also incorporates a positioning mechanism that ensures consistent articulatory placement through real-time signal correlation with preset reference positions before each pronunciation.

\begin{figure}[t]
\centering
\includegraphics[width=0.7\columnwidth]{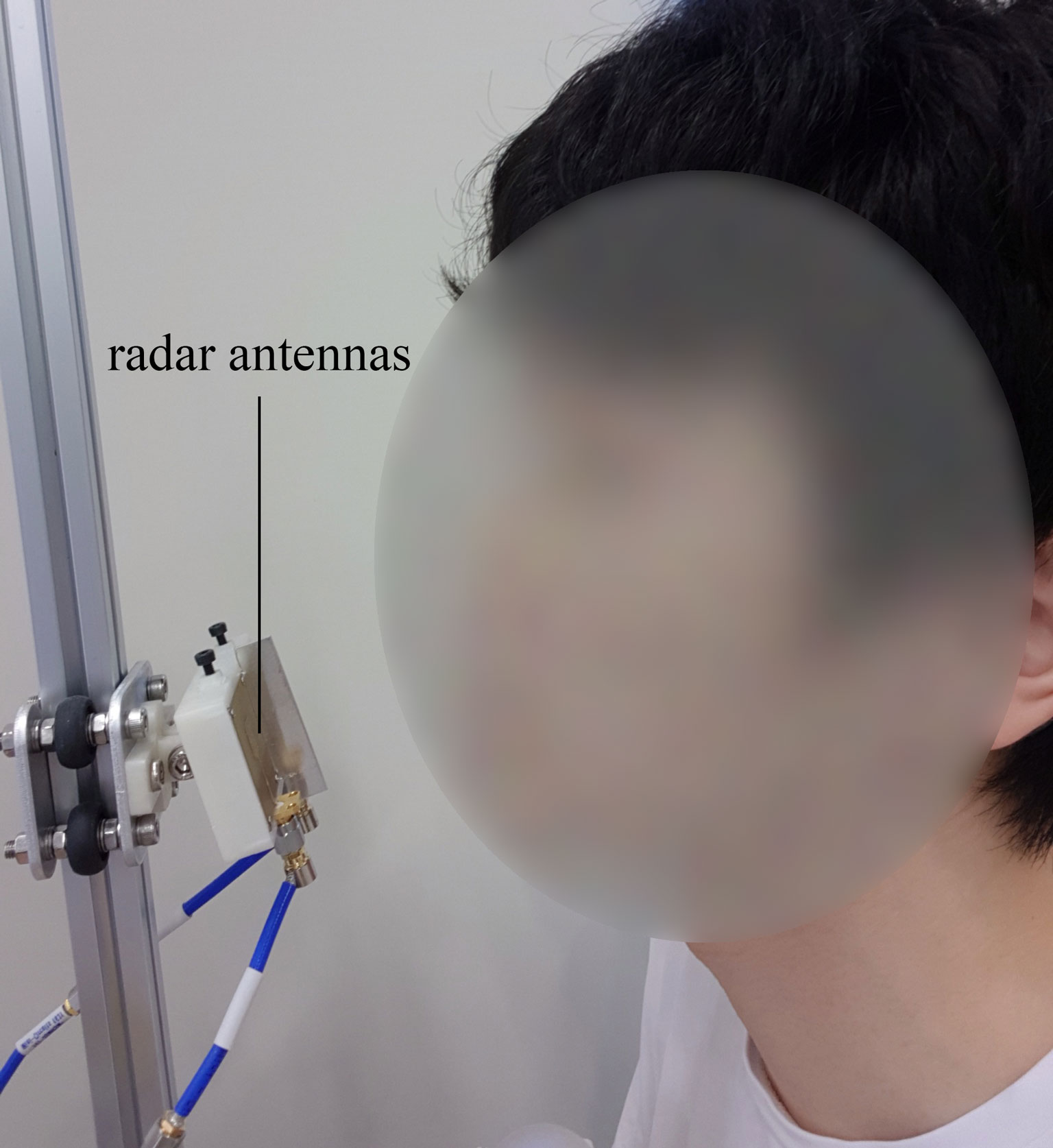}
\caption{Our IR-UWB radar SSR data collection setup.}
\label{fig:setup}
\end{figure}

Data were collected from a single participant (male, 30 years old) who was a native Korean speaker with more than 20 years of English education and no known speech or hearing impairments.
Fig. \ref{fig:setup} shows our experimental setup for contactless IR-UWB radar-based silent speech data collection.
The participant was positioned in front of the radar antennas, with lips positioned 5--10 cm away from the antennas.

We used 50 phonetically balanced English words from \cite{Haskins1949} as speech stimuli.
The participant pronounced each word across 20 sessions.
This resulted in a total of 1,000 samples (50 words × 20 repetitions) for the dataset.

\subsection{Data Representation}

Each IR-UWB radar measurement comprises a frame with $N$-dimensional amplitudes, where $N = 256$ in this research.
After pronouncing each speech stimulus, a complete frame set capturing articulatory movements in $M$ frames is acquired at approximately 100 frames per second in this study.
The frame set forms an $M \times N$ matrix where $M$ represents the number of slow-time indices (sequential radar measurements) and $N$ represents the number of fast-time indices (range bins).
This matrix structure enables monitoring of articulatory movements through temporal and spatial radar reflection changes.

The raw radar data undergoes three light preprocessing steps to enhance signal quality and reduce data dimensionality.
First, clutter removal is performed as a standard radar processing procedure, estimating and removing static background reflections using an exponential moving average with adaptation rate $\alpha = 0.95$ \cite{Lee23:IR-UWB}.
Subsequently, fast-time indices 1--100 are selected from the $M \times N$ frame set matrix to focus on the spatial regions most relevant for articulatory movement detection.
Finally, DC offset removal is conducted by subtracting the mean value from each individual frame.
This light preprocessing pipeline transforms the original $M \times 256$ radar measurements into a focused $M \times 100$ matrix representation, preserving essential speech articulation patterns while reducing the input dimensionality for the neural network.

\subsection{Attention-Enhanced TCN Architecture}

\begin{figure}[t]
\centering
\includegraphics[width=0.35\columnwidth]{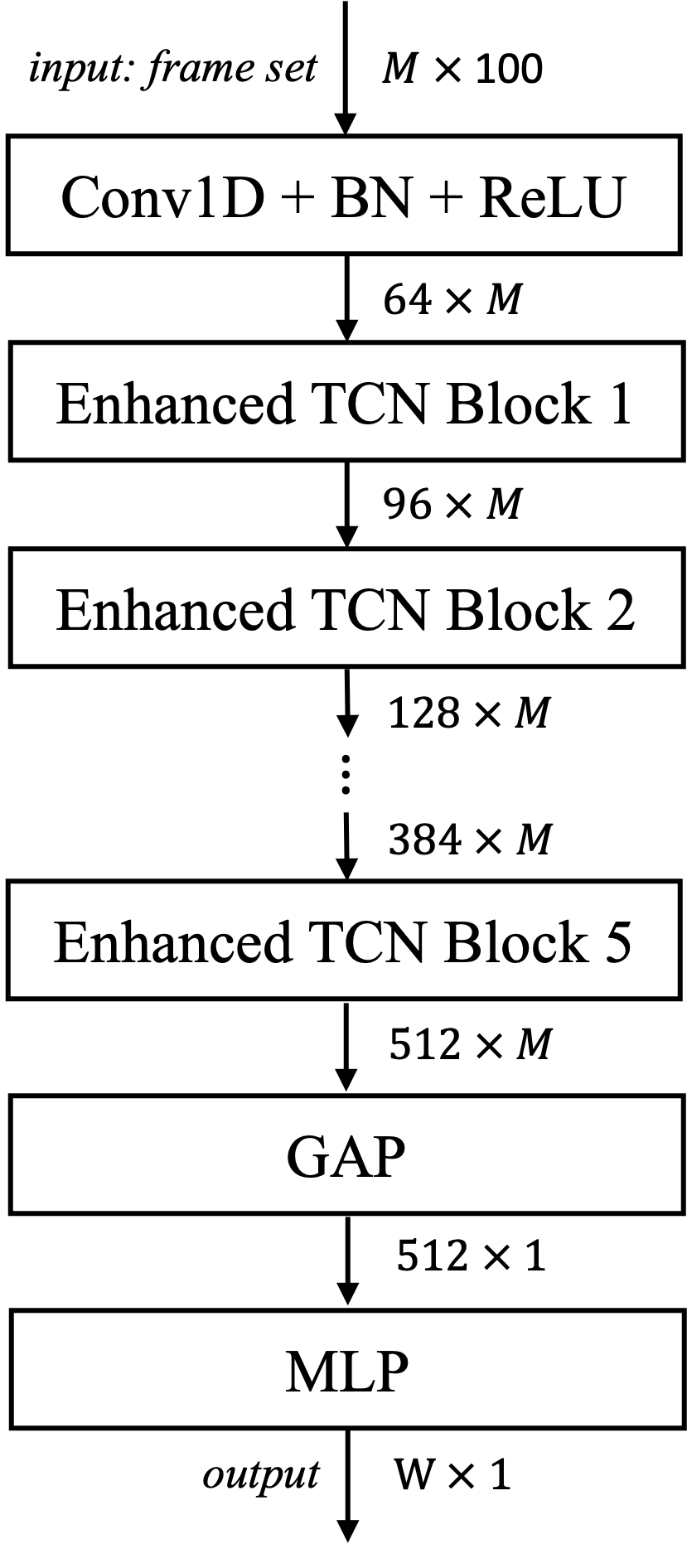}
\caption{Network topology of the attention-enhanced TCN for IR-UWB radar-based SSR.}
\label{fig:architecture}
\end{figure}

\begin{figure}[t]
\centering
\includegraphics[width=0.35\columnwidth]{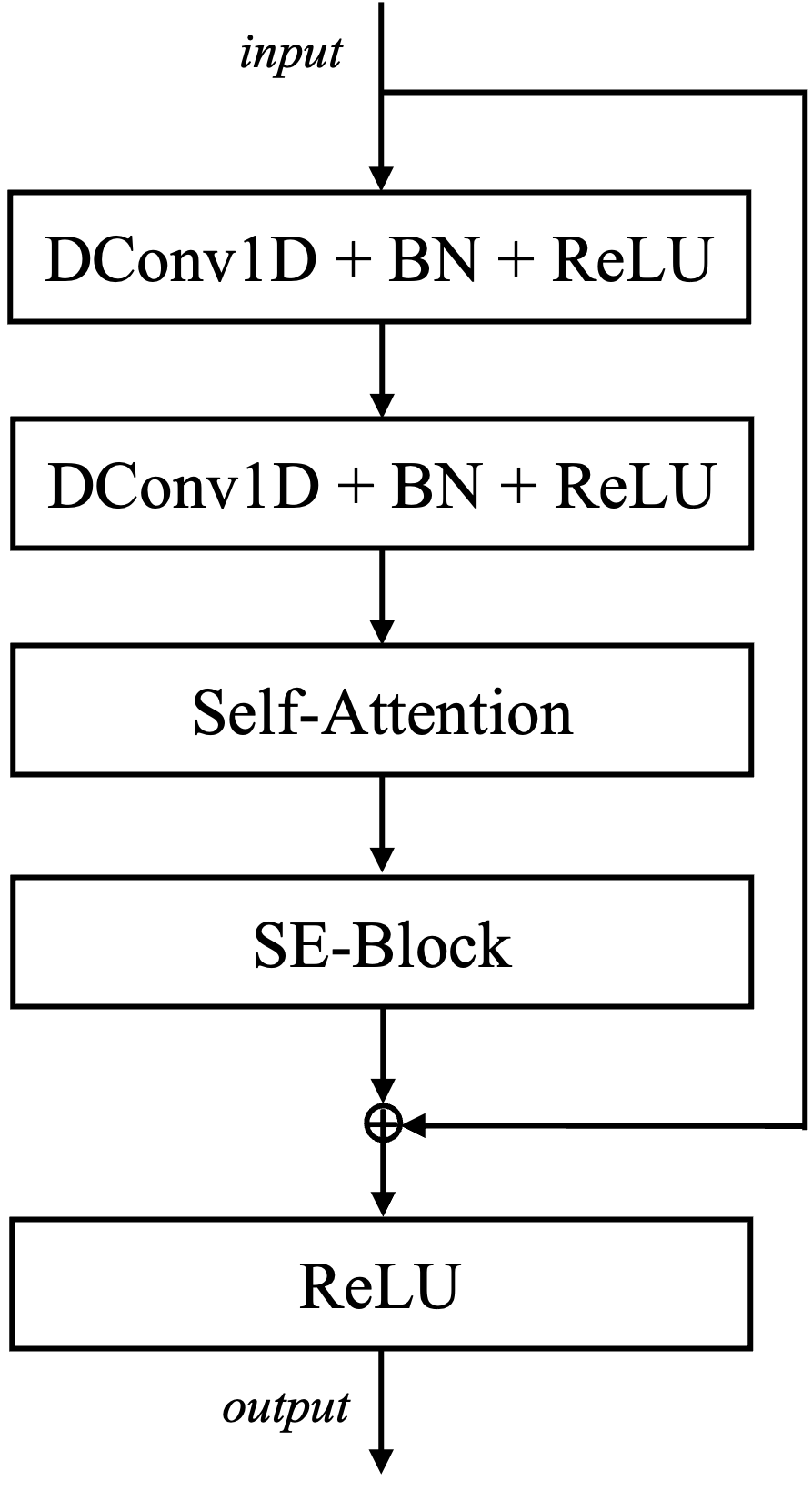}
\caption{The enhanced TCN block of the attention-enhanced architecture.}
\label{fig:tcn_block}
\end{figure}

Our approach employs attention-enhanced temporal convolutional networks that process the lightly preprocessed radar frame sets through a hierarchical architecture designed for end-to-end IR-UWB radar-based SSR. 
The architecture builds upon the temporal convolutional network framework \cite{Bai2018:TCN}, integrating self-attention \cite{Vaswani2017:Attention} and squeeze-and-excitation \cite{Hu2018:SE} blocks to enhance feature representation learning.

Fig. \ref{fig:architecture} presents the overall network topology, which consists of three main components operating in sequence.
First, an input processing layer handles variable-length sequences and performs initial feature transformation.
Second, a stack of five enhanced TCN blocks with integrated attention mechanisms progressively extract hierarchical temporal features.
Finally, a classification head with global pooling produces the final word classification.

The network employs a hierarchical design comprising an input layer followed by five enhanced TCN blocks, progressively expanding both feature representation capacity and temporal receptive field.
The channel dimensions progress as $C_i \in \{64, 96, 128, 256, 384, 512\}$ for $i = 0, ..., 5$, where $C_0 = 64$ is the output of the input processing layer, while dilation rates $d_i = 2^{i-1}$ for $i = 1, ..., 5$ increase exponentially in the TCN blocks to capture long-range temporal dependencies.
This progressive architecture enables the model to capture both fine-grained local patterns and long-range temporal dependencies essential for speech recognition from radar signals.

The input processing layer receives the lightly preprocessed frame set $\mathbf{x}$ of dimensions $M \times 100$, where $M$ corresponds to slow-time indices varying with utterance duration and 100 represents the selected fast-time indices, and applies initial feature transformation:
\begin{equation}
\mathbf{h}_0 = \text{ReLU}(\text{BN}(\text{Conv1D}(\mathbf{x})))
\end{equation}
where $\text{Conv1D}$ denotes one-dimensional convolution with kernel size of 3 and padding of 1, followed by batch normalization and ReLU activation, transforming from 100 to 64 channels while preserving temporal dimension.

Fig. \ref{fig:tcn_block} shows the structure of the enhanced TCN block, which serves as the core building unit of our architecture.
Each enhanced TCN block consists of four key components: (1) two cascaded dilated convolution blocks for temporal feature extraction, (2) self-attention mechanism for capturing long-range dependencies, (3) squeeze-and-excitation module for channel-wise recalibration, and (4) residual connection for gradient flow.

First, each block processes the previous output $\mathbf{h}_{i-1}$ through two cascaded dilated convolution blocks.
We define a dilated convolution block as:
\begin{equation}
\text{DCBlock}_{d}(x) = \text{ReLU}(\text{BN}(\text{DConv1D}_{d}(x)))
\end{equation}
where $\text{DConv1D}_{d}$ denotes one-dimensional dilated convolution with dilation rate $d$ and appropriate padding to maintain temporal dimension, followed by batch normalization and ReLU activation.
Each convolution uses kernel size 3, and dropout with rate 0.25 is applied after each ReLU activation for regularization.
The complete feature extraction within each TCN block is:
\begin{equation}
\mathbf{f}_i = \text{DCBlock}_{d_i}(\text{DCBlock}_{d_i}(\mathbf{h}_{i-1}))
\end{equation}
with dilation rate $d_i = 2^{i-1}$.
The complete feature extraction transforms $\mathbf{h}_{i-1} \in \mathbb{R}^{C_{i-1} \times M}$ to $\mathbf{f}_i \in \mathbb{R}^{C_i \times M}$ through two cascaded dilated convolutions, where the channel transformation occurs in the first convolution.

Next, the enhanced features $\mathbf{f}_i$ are processed through attention mechanisms.
The self-attention mechanism computes scaled dot-product attention:
\begin{equation}
\text{Attention}(Q,K,V) = \text{softmax}\left(\frac{QK^T}{\sqrt{d_k}}\right)V
\end{equation}
where $Q = \mathbf{f}_i W_Q$, $K = \mathbf{f}_i W_K$, $V = \mathbf{f}_i W_V$ are the query, key, and value projections with $d_k = C_i/8$. Here, $C_i$ denotes the channel dimension at level $i$ (e.g., $C_3 = 256$).
To ensure training stability, the attention output is gated by a learnable scalar parameter $\gamma$ initialized to zero, resulting in:
\begin{equation}
\text{SelfAttn}(\mathbf{f}_i) = \gamma \cdot \text{Attention}(Q,K,V) + \mathbf{f}_i
\end{equation}
This allows the network to gradually learn the contribution of attention during training.

Subsequently, the squeeze-and-excitation block applies channel-wise recalibration:
\begin{equation}
\text{SE}(\mathbf{x}) = \mathbf{x} \cdot \sigma(\text{FC}_2(\text{ReLU}(\text{FC}_1(\text{GAP}(\mathbf{x})))))
\end{equation}
where $\text{GAP}$ denotes global average pooling, $\text{FC}_1$ reduces channels by factor of 16, and $\text{FC}_2$ restores original dimensionality with sigmoid activation $\sigma$.

Finally, the complete enhanced TCN block combines all these components:
\begin{equation}
\mathbf{h}_i = \text{ReLU}(\text{SE}(\text{SelfAttn}(\mathbf{f}_i)) + \text{Residual}(\mathbf{h}_{i-1}))
\end{equation}
where residual connections \cite{He2016:ResNet} employ 1×1 convolutions when dimension matching is required, facilitating gradient flow across the deep network.

The classification head maps the final features $\mathbf{h}_5$ into word-level predictions using global average pooling followed by a two-layer perceptron:
\begin{equation}
\text{output} = \text{MLP}(\text{GAP}(\mathbf{h}_5))
\end{equation}
Here, MLP first reduces dimensionality from 512 to 256 with batch normalization, ReLU activation, and dropout rate 0.25, and then produces logits from 256 to $W=50$ word classes.

\section{Experiments}

This study evaluates the attention-enhanced TCN using a leave-one-session-out cross-validation approach to ensure robust performance assessment.
The experimental design uses 20 sessions total, where each fold designates one session as the test set, while the remaining 19 sessions are split into training (16 sessions) and validation (3 sessions) sets.
The validation sessions are randomly selected for each fold using a fixed seed to ensure reproducibility while maintaining diversity in training configurations across folds.

Training employs the AdamW optimizer with an initial learning rate of 0.0008, weight decay of 1e-4, and cosine annealing with warm-up scheduling.
The batch size is set to 32 with label smoothing ($\alpha = 0.15$) and early stopping (patience = 18 epochs) based on validation loss.
Gradient clipping with maximum norm 1.0 is applied to ensure training stability.
Training typically requires 70--140 epochs per fold, with early stopping preventing overfitting.

For comparison, the proposed approach is evaluated against the conventional method that employs hand-crafted features followed by deep neural network--hidden Markov model (DNN--HMM), as demonstrated in prior IR-UWB radar-based SSR research \cite{Lee23:IR-UWB}.

Classification accuracy serves as the primary evaluation metric, computed as the percentage of correctly classified word instances.
Results are presented as both individual fold performance and average performance across all 20 folds.

\section{Results}

\begin{table}[t]
\centering
\caption{Leave-one-session-out cross-validation results for 50-word English vocabulary recognition using IR-UWB radar.}
\label{tab:results}
\begin{tabular}{ccc}
\hline
Test & Baseline & Proposed \\
Session & (hand-crafted + DNN--HMM) & (attention-enhanced TCN) \\
\hline
1 & 60\% & 88\% \\
2 & 74\% & 92\% \\
3 & 78\% & 94\% \\
4 & 76\% & 88\% \\
5 & 86\% & 92\% \\
6 & 64\% & 90\% \\
7 & 94\% & 92\% \\
8 & 82\% & 92\% \\
9 & 74\% & 90\% \\
10 & 74\% & 92\% \\
11 & 60\% & 84\% \\
12 & 58\% & 98\% \\
13 & 74\% & 88\% \\
14 & 84\% & 94\% \\
15 & 76\% & 92\% \\
16 & 56\% & 92\% \\
17 & 70\% & 92\% \\
18 & 78\% & 86\% \\
19 & 74\% & 90\% \\
20 & 88\% & 96\% \\
\hline
\textbf{Average} & \textbf{74.0\%} & \textbf{91.1\%} \\
\hline
\end{tabular}
\end{table}

Table \ref{tab:results} presents the comprehensive results of our leave-one-session-out cross-validation evaluation.
Our attention-enhanced TCN achieves an average test accuracy of 91.1\% across 20 folds, significantly outperforming the conventional hand-crafted features with DNN--HMM approach, which achieves 74.0\% average accuracy.
This demonstrates the substantial benefit of end-to-end learning for IR-UWB radar-based SSR.

Our attention-enhanced TCN demonstrates superior consistency compared to the baseline approach.
The standard deviation of our method (3.3\%) is significantly lower than the hand-crafted feature approach (10.3\%), indicating more stable performance across different test sessions.
This consistency is crucial for practical SSR applications where reliable performance is essential.
Even the lowest performance (84\% in Fold 11) substantially exceeds the baseline method's average performance.

The substantial improvement can be attributed to several key factors.
By learning features directly from radar signals, our approach eliminates the bottleneck of hand-crafted feature design and adapts representations specifically for the classification task.
The multi-level TCN hierarchy with increasing dilation rates forms the backbone of the model, enabling the capture of both local articulatory patterns and long-range temporal dependencies essential for word recognition.
On top of this backbone, the integration of self-attention and SE-blocks further enhances performance by allowing the model to focus on discriminative temporal segments and channel responses, effectively handling the complexity of radar-based articulatory movement detection.

\section{Conclusion}

This paper presents an attention-enhanced temporal convolutional network architecture for IR-UWB radar-based silent speech recognition.
The results demonstrate that this approach achieves significant improvement in recognition accuracy compared to the conventional method, with an average test accuracy of 91.1\% on a 50-word classification task.
This study demonstrates the potential of end-to-end learning for IR-UWB radar-based SSR and paves the way for practical applications.

\bibliographystyle{IEEEtran}
\bibliography{mybibfile, IUS_publications}

\vspace{12pt}

\end{document}